\documentclass{elsart}

\def\ii{\'{\i }}

\usepackage{graphicx,amssymb}
\journal{Physics Letters B}

\def\maiorsim{\smash{\mathop{>}\limits_{\raise3pt\hbox{$\sim$}}}}

\def\infinito{\smash{\mathop{\longrightarrow}\limits_{\raise3pt\hbox{$_{\sqrt s \to \infty}$}}}}

\def\doisdeltas{\smash{\mathop{\longrightarrow}\limits_{\raise3pt\hbox{$_{\Delta >\Delta_0 (y_b)}$}}}}

\begin{document}
\begin{frontmatter}

\title{Forward-backward rapidity correlations in a two-step scenario}%
\author{P. Brogueira}

\address{Departamento de F\ii sica, IST, Av. Rovisco Pais, 1049-001 Lisboa, Portugal}

\author{J. Dias de Deus\corauthref{JDD}}
\corauth[JDD]{Corresponding author.}

\ead{jdd@fisica.ist.utl.pt}

\address{CENTRA, Departamento de F\ii sica, IST, Av. Rovisco Pais, 1049-001 Lisboa, Portugal} 

\begin{abstract}
We argue that two-step models, like String and Glasma models, with creation first of sources extended in rapidity that after locally decay into particles, lead to long range forward-backward correlations due to fluctuations in the number or the colour of the sources, and to a similar expression for the {\it F-B} correlation parameter $b$.

In the simplest String percolation scenario, $b$ increases with centrality or the number of participating nucleons. However, asymptotically it decreases with energy. This is different from the Glasma model where $b$ increases with both, centrality and energy.
\end{abstract}

\end{frontmatter}


The study of particle correlations and fluctuations can give an important contribution to the understanding of the behaviour of matter at high density in high energy heavy ion collisions. Interesting experimental results have been recentely presented, [1,2,3,4], as well as various theoretical developments (see [5] and references therein). In this paper we discuss forward-backward rapidity correlations in the general framework of models with particle production occuring in a two-step scenario.

In models where particle production occurs in two steps, formation first of a rapidity structure reflecting the longitudinal distribution of the colour fields (as it happens with the Glasma [6] or with the dual string model [7]), followed by local emission of particles, the presence of long distance rapidity forward-backward correlations is, in a sense, unavoidable. Even if particle emission from the created sources is totally uncorrelated, as usually assumed, long range correlations occur in the process of averaging over fluctuations in the colour or number of the emitting sources. In multi-collision processes, as nucleus-nucleus collisions, such fluctuations are naturally expected [8,9].

In the simplest model, all the extended in rapidity sources are equal, and fluctuate in their number or colour $N$, and $n$ particles are locally produced, with the result [9]:
$$
\langle n\rangle =\bar N \bar n \ ,\eqno(1)
$$
where $\langle n\rangle$ is the average multiplicity, and $\bar n$ the single source average multiplicity, and
$$
D^2/\langle n\rangle^2 = {\overline{N^2} -{\overline N}^2 \over \bar N^2} + {1\over \langle n\rangle} {d^2\over \bar n} \ ,\eqno(2)
$$
where $D^2 \equiv \langle n^2\rangle - \langle n\rangle^2$ and $d^2 \equiv \overline{n^2} -{\overline n}^2$. The first term in the right hand side of (2) represents the rapidity long range correlation, and the second term the local short-range one. It is clear that the dynamics associated to the initial conditions of the collision is controled by the first (long range correlation) term in (2). The second term controls local production (resonances, short range correlations, etc.). For simplicity we shall assume that the local, single source particle distribution is a Poisson, which means
$$
d^2/\bar n =1 \ .\eqno(3)
$$

From (2) and (3) we obtain
$$
D^2 = {\langle n\rangle^2\over K} + \langle n\rangle \ ,\eqno(4)
$$
with
$$
K\equiv {\overline N^2 \over \overline{N^2}- {\overline N}^2} \ . \eqno(5)
$$
It should be noticed that the quantities $\bar n, \langle n\rangle, \bar N$ and $K$ are, in principle, dependent on the width and position of the rapidity window under study. We shall concentrate on symmetrical, $AA$, collisions.

If we consider a window in the forward direction (or backward direction) we obtain
$$
\langle n_f\rangle = \langle n_b\rangle = \bar N \bar n \ ,\eqno(6)
$$
and
$$
D^2_{ff} = {\langle n_f\rangle^2 \over K} + \langle n_f\rangle \ .\eqno(7)
$$
If we consider next two symmetrically placed windows, one forward and one backward, separeted by a rapidity gap, such that the short range contribution is practically eliminated, we have
$$
D^2_{bf} = {\langle n_f\rangle^2 \over K} \ .\eqno(8)
$$

The correlation parameter $b$, in the relation $\langle n_b\rangle_F = a+b n_F$, is given by
$$
b\equiv D^2_{bf}/D^2_{ff} \ ,\eqno(9)
$$
and we obtain
$$
b={1\over 1+K/\langle n_f\rangle} \ .\eqno(10)
$$
This result was previously obtained in two different models, [10] and [11], having in common the presence of sources extending along the rapidity axis, and fluctuations in their number or overall colour.

In the Glasma approach, [11], the long range correlation piece behaves as $1/\alpha_s$, while the short range correlation one behaves as $\alpha_s$, such that $K/\langle n_f\rangle \sim \alpha^2_s$, $\alpha_s$ being the QCD coupling constant. The {\it F-B} correlation parameter $b$ increases with centrality or the number of participating nucleons, $N_{part}$, the same happenning with the increase of the energy.

In the String approach we do not have such straightforward result: one has to analize the quantities $\langle n_f\rangle$ and $K$. This is what we will do next.

It is well known that dual string models cannot reproduce correctely the height of the rapitity plateau [12] and the $N_{part}$ dependence of $dn/dy$ [13] without inclusion of string fusion, percolation, or any other multiplicity suppression mechanism.

In string percolation, the relevant parameter is the transverse density of strings, $\eta$,

$$
\eta \equiv \left( {r\over R}\right)^2 N(\sqrt s) \ ,\eqno(11)
$$
where $r$ is the string radius, $R$ the transverse radius of the overlap region and $N(\sqrt s)$ the number of strings at a given energy. Making use of the standard relations in high energy interactions with nuclei, [12,13],

\begin{figure}
\begin{center}
\includegraphics*[width=10cm]{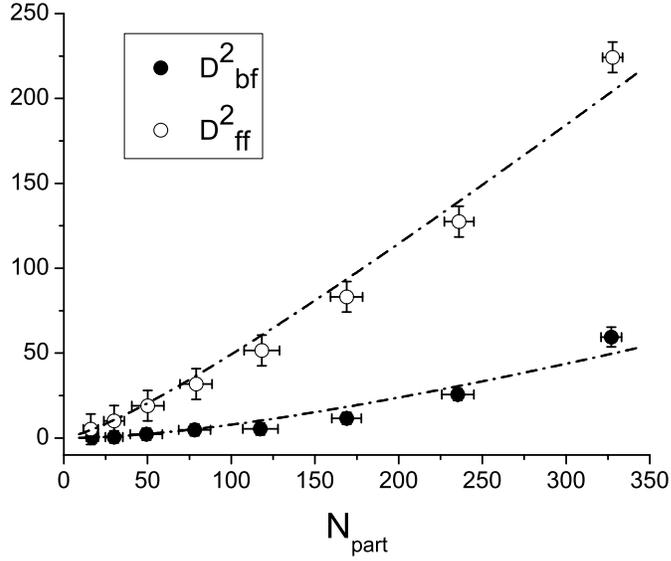}      
\end{center}
\caption{$D^2_{ff}$, Eq.(7), and $D^2_{bf}$, Eq.(8), as functions of $N_{part}$, by making use of (16) and (18). Data are from [3]. Errors were estimated from the figures of [3].}
\end{figure}

$$
N\simeq N_p \left( {N_{part}\over 2}\right)^{4/3} \ ,\ R\simeq R_p \left( {N_{part}\over 2}\right)^{1/3} \ ,\eqno(12)
$$
where $N_p$ is the number of strings in $pp$ collisions and $R_p$ the nucleon radius, we obtain
$$
\eta \equiv \left( {r \over R_p}\right)^2 N_p (\sqrt s )\left( {N_{part}\over 2}\right)^{2/3} \ ,\eqno(13)
$$

For the moment, as we are only interested in the $N_{part}$ dependence, we write, from (13),
$$
\eta =A N_{part}^{2/3} \ ,\eqno(14)
$$

\noindent $A$ being a parameter. From recent and not so recent work [14], it is known that for $N_{part} \simeq 350$, central Au-Au collisions, $\eta \simeq 3$, which means $A\simeq 0.06$. We fix this value for $A$.

The most important consequence of percolation is the presence of the colour reduction factor $F(\eta )$,[15],
$$
F(\eta ) \equiv \sqrt{1-e^{-\eta} \over \eta}\ ,\eqno(15)
$$
such that (1), for the forward region, can be rewritten in the form
$$
\langle n_f\rangle =F (A N_{part}^{2/3} )BN_{part}^{4/3} \ ,\eqno(16)
$$

\noindent $B$ being a free parameter. Note that asymptotically $F(\eta) \to \eta^{-1/2}$.

Regarding the inverse of the normalized variance of the source distribution, $K$, it is known to increase with $N_{part}$, for $N_{part}\maiorsim 30$ [2,4]. We simply assume that string correlations, for large density, are proportional to the density $\eta :\bar N \sim \eta , (\overline{ N^2} - \bar N^2 -\bar N )\sim \eta$, ..., such that 

\begin{figure}
\begin{center}
\includegraphics*[width=11cm]{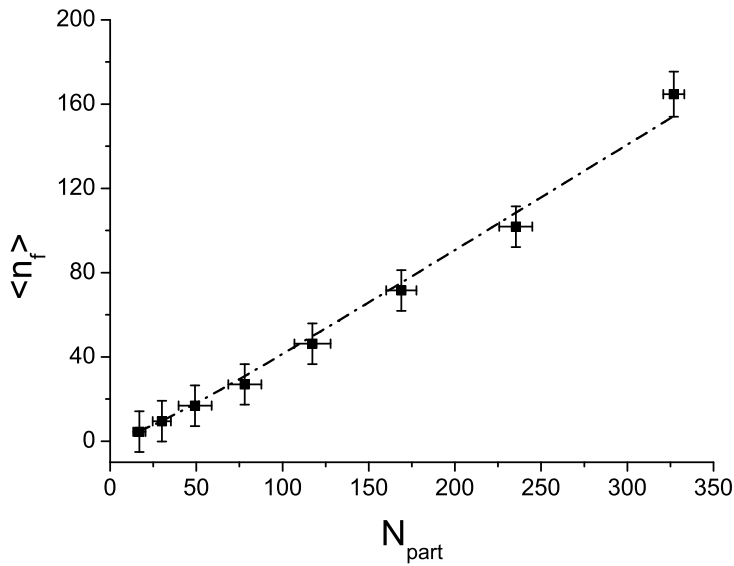}      
\end{center}
\caption{a) $\langle n_F \rangle$ as a function of the number of participants $N_{part}$. The curve is from (16) $B=0.119 \pm 0.006$.}
\end{figure}

$$
K\sim \eta \ ,\eqno(17)
$$
and we write, see (14),
$$
K= C N_{part}^{2/3} \ ,\eqno(18)
$$

\noindent $C$ being a new parameter. 

Note that relations (17) and (18) in the framework of percolation should not apply in the small $\eta$ region: the $K(\eta)$ should have a minimum and even increase with $\eta$, as $\eta \to 0$, as seen in $pp$ collisions [16]. We can take that behaviour into account, but it does not change the final result as RHIC data are not sensitive to small $N_{part}$.

Making use of (18) and (16) with (15) one sees that, for large $N_{part}$,
$$
K/\langle n_f\rangle \rightarrow N_{part}^{-1/3} \ ,\eqno(19)
$$
which means that the correlation parameter $b$ increases with centrality, or the number of participants, as in [11]. In the no-percolation limit, $F(\eta)\equiv 1$, $K/\langle n_f\rangle \to N_{part}^{-2/3}$, which means that $b$ shows a faster increase with centrality.

\begin{figure}
\begin{center}
\includegraphics*[width=11cm]{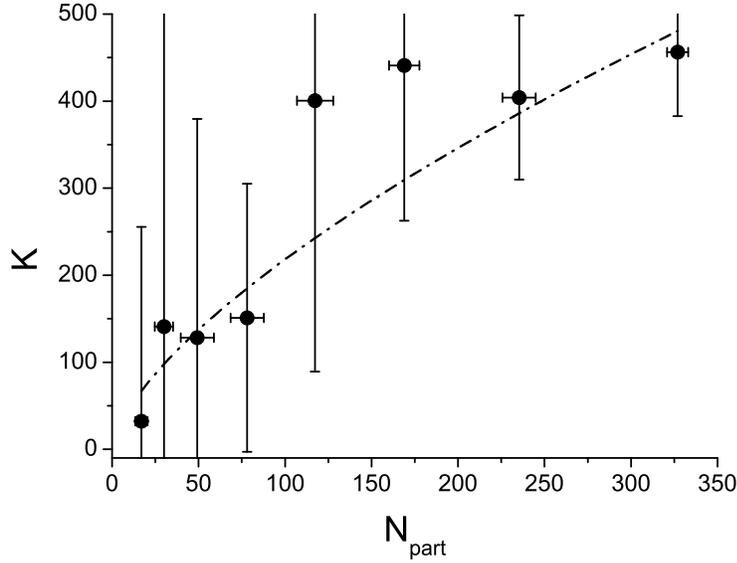}      
\end{center}
\caption{$K$ as a function of $N_{part}$. The curve is from (18) with $C=10.1 \pm 1.3$.}
\end{figure}

Recentely, the STAR collaboration at RHIC, [3], has presented results for $D^2_{ff}$ and $D^2_{bf}$, shown in Fig.1, as a function of $N_{part}$ in $Au-Au$ at $\sqrt s =200$GeV. Making use of the experimental points and of equations (7), for $D^2_{ff}$, and (8) for $D^2_{bf}$, we have extracted the "experimental" points for $\langle n_f\rangle$ (Fig.2) and $K$ (Fig.3), and fitted them with equations (16), for $\langle n_f\rangle$, and (18), for $K$ (see the curves in Figs. 1,2 and 3). Note that $K (N_{part})$ in Fig.3 is not very much constrained by data, in  particular in the low $N_{part}$ region, as mentioned above.

It is interesting to ask another question: what happens to the {\it F-B} correlations at fixed $N_{part}$, with increasing energy? The answer is, see (13),
$$
\langle n_f\rangle =F(A' N_p(\sqrt s)) B' N_p (\sqrt s) \ ,\eqno(20)
$$
and 
$$
K=C' N_p (\sqrt s) \ ,\eqno(21)
$$

\noindent $A'$, $B'$ and $C'$ being parameters, such that, asymptotically,
$$
K/\langle n_f\rangle \infinito  N_p (\sqrt s)^{1/2} \ , \eqno(22)
$$
which means that the {\it F-B} correlation parameter $b$ {\bf decreases} as $\sqrt s$ goes to infinity. In the absence of percolation, $F(\eta)\equiv 1$, $b$ is essentially constant. In the Glasma model, [11], $b$ increases with the energy. We may thus have here a possibility of distinguish the Glasma model from the String Percolation model. As in moving from RHIC to LHC the number of strings in supposed to increase by a factor around 2 (see, for instance, the second reference in [14]), we expect at LHC a reduction of $b$, for central collisions and similar phase-space, of the order of 25\% in comparison with RHIC results. However it should be mentioned that this result strongly depends on the validity of (17) (see [16]). If $K \sim \eta^a$, with $a<1/2$, then $b$ also increases with energy in the percolation scenario.

{\it Acknowledgments}

We would like to thank Carlos Pajares and Brijesh Srivastava for discussions and Jeff Mitchell and Brijesh Srivastava for information on RHIC data. 

{\it References}

\begin{enumerate}
\item B.B. Back et al. (PHOBOS Coll.), Phys. Rev. C74 (2006) 011901.
\item M. Rybczynski et al. (NA49 Coll.), nucl-ex/0409009 (2004).
\item T.J. Tarnowsky (STAR Coll.), nucl-ex/0606018, Proc. 22nd Winter Workshop on Nuclear Dynamics (2006).
\item J.T. Mitchell (PHENIX Coll.), nucl-ex/0511033 (2005).
\item S. Haussler, M. Abdel-Aziz, M. Bleicher, nucl-th/0608021.
\item A. Kovner, L.D. McLerran and H. Weigert, Phys. Rev. D52 (1995) 6231; T. Lappi and L.D. McLerran, Nucl. Phys. A772 (2006) 200.
\item N.S. Amelin, N. Armesto, M.A. Braun, E.G. Ferreiro and C. Pajares, Phys. Rev. Lett. 73 (1994) 2813.
\item A. Capella and A. Krzywicki, Phys. Rev. D18 (1978) 4120; A. Capella and J. Tran Thanh Van, Phys. Rev. D (1984) 2512.
\item J. Dias de Deus, C. Pajares and C. Salgado, Phys. Lett. B407 (1997) 335.
\item M.A. Braun, C. Pajares and V.V. Vecherin, Phys. Lett. B493 (2000) 54.
\item N. Armesto, L. McLerran and C. Pajares, hep-ph/0607345.
\item N. Armesto and C. Pajares, Int. J. Mod. Phys. A15 (2000) 2019.
\item J. Dias de Deus and R. Ugoccioni, Phys. Lett. B491 (2000) 253; Phys. Lett. B494 (2000) 53.
\item B.K. Srivastava, R.P. Scharengerg, T.J. Tarnowsky, nucl-ex/0606019 (2006); N. Armesto, M.A. Braun, E.G. Ferreiro, C. Pajares, Phys. Rev. Lett. 77 (1996), 3736.
\item M.A. Braun, F. Del Moral and C. Pajares, Phys. Rev. C65 (2002) 02490.
\item J. Dias de Deus, E.G. Ferreiro, C. Pajares and R. Ugoccioni, Phys. Lett. B601 (2004) 125.
\end{enumerate}
\end{document}